\documentclass[a4paper]{jpconf}
\usepackage{graphicx}

\usepackage{setspace} 
\usepackage{paralist} 
\usepackage[retain-zero-exponent, expproduct=tightcdot]{siunitx}

\begin{document}
\title{High-dynamic-range water window ptychography}


\author{Max Rose$^1$, Dmitry Dzhigaev$^{1}$, Tobias Senkbeil$^2$, Andreas R. von Gundlach$^2$, Susan Stuhr$^2$, Christoph Rumancev$^2$, Ilya Besedin$^1$, Petr Skopintsev$^1$, Jens Viefhaus$^1$, Axel Rosenhahn$^{2}$ and Ivan A. Vartanyants$^{1,3}$}
\address{
$^{1}$Deutsches Elektronen-Synchrotron DESY, Hamburg, 22607, Germany\\
$^{2}$Analytical Chemistry - Biointerfaces, Ruhr-University Bochum, Bochum, 44780, Germany\\
$^{3}$MEPhI (Moscow Engineering Physics Institute), Moscow, 115409, Russia}


\ead{max.rose@desy.de}


\begin{abstract}
Ptychographic imaging with soft X-rays, especially in the water window energy range, suffers from limited detector dynamic range that directly influences the maximum spatial resolution achievable. High-dynamic-range data can be obtained by multiple exposures. By this approach we have increased the dynamic range of a ptychographic data set by a factor of \num{76} and obtained diffraction signal till the corners of the detector. The real space half period resolution was improved from \SI{50}{\nano\meter} for the single exposure data to \SI{18}{\nano\meter} for the high-dynamic-range data.

\end{abstract}

\section{Introduction}
Ptychography is a scanning coherent diffractive imaging (CDI) method for extended samples \cite{Rodenburg2007a, Thibault2008a}.
It provides a self consistent solution to the phase problem in CDI by overlapping illuminations on the measured object. In addition, it can also recover the complex valued illumination function.
Especially in the field of X-ray microscopy (XRM), which lacks efficient and high quality imaging optics, ptychography is a powerful lens-less imaging technique
\cite{Murphy2013}.
Ptychography also decouples the illumination beam size from the final resolution of the picture, which is advantageous over scanning transmission X-ray microscopy (STXM).

The so-called water window energy range from
\SI{280}{\electronvolt}
to
\SI{530}{\electronvolt}
contains the absorption edges of oxygen, nitrogen and carbon and proteins and carbon rich material are visible with high contrast against the aqueous environment of biological samples
\cite{Schneider1998a}.
This makes the water window unique and especially well suited for high resolution XRM
\cite{Larabell2010a}
and subsequently also for X-ray ptychography
\cite{Giewekemeyer2011b}
of unstained biological samples.

In the soft X-ray range silicon based integrating pixel detectors are common devices for measuring diffraction patterns. Unfortunately, they have limitations in the context of coherent diffractive imaging. Due to the integrating nature of the detector and electronic imperfections, noise and dark-current and the maximum photon number per pixel
(\SI{898}{photons \per pixel})
the effective dynamic range (DR) of the detector is limited.
This is a severe limitation for CDI and water window ptychography of weakly scattering samples due to the large DR of diffraction patterns
\cite{Rose2015a}. Different approaches exist to circumvent this problem. Among them are
\begin{inparaenum}[i)]
\item pixel capacity enlargement
\item active pixel gain switching \cite{Wunderer2014a} and
\item multiple exposure merging \cite{Takahashi2011a}.
\end{inparaenum}
Pixel capacity enlargement requires pixel size enlargement. This is often not tolerable in diffractive imaging due to the required oversampling criterion
\cite{Miao1999a}.
Detectors with active pixel gain switching are being under development but they are currently not available to a broader community.

\section{Experiment}
Here we present a ptychographic experiment on a known test sample  (NTT-AT, XRESO-50HC) in order to characterize the detailed beam properties on a high resolution scale. We measured multiple exposures in our experiment to obtain high-dynamic-range (HDR) diffraction patterns by the combination of multiple data sets. The experiment was performed at the XUV Beamline P04 at PETRA III \cite{Viefhaus2013a}.

\begin{figure}[htbp]
\centering
\includegraphics[width=0.9\linewidth]{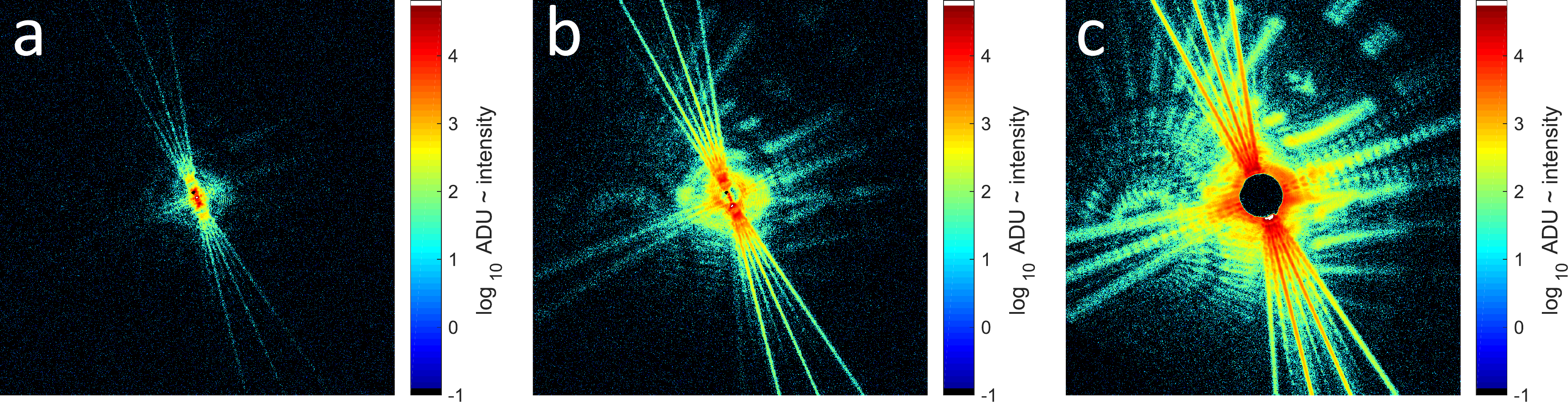}
\caption{Siemens star far-field diffraction patterns of different exposure time. a) Short exposure $t_\textrm{noBS} = \SI{0.15}{\sec}$, b) long exposure $t_\textrm{BS1} = 15 \cdot t_\textrm{noBS}$ with semi-transparent beamstop and c) long exposure $t_\textrm{BS2} = 150 \cdot t_\textrm{noBS}$ with opaque beam stop.}
\label{fig:diffPatterns}
\end{figure}

The diffraction patterns from the Siemens star test sample were measured with an ANDOR
\SI{16}{bit}
detector (DODX436-BN) at
\SI{500}{\electronvolt}
with the detector at a distance of
$z = \SI{26.2}{\centi\meter}$.
The incident beam on the sample was originating from a
\SI{3}{\micro\meter}
pinhole at a distance of
\SI{1}{\milli\meter}. We scanned our sample on a equidistant rectangular grid
(\num{8 x 9} positions) with a step size of
\SI{1.2}{\micro\meter}.
We used the following strategy for the measurement: The first scan was obtained without beam stop, the second scan with a semi-transparent beam stop (BS1) of
\SI{1.6}{\milli\meter}
diameter and the third scan with an opaque beam stop (BS2) of
\SI{3.4}{\milli\meter}.
The resulting diffraction patterns (background and dark current corrected) in analog-digital units (\SI{73}{ADU\per photon}) are shown on a logarithmic scale in
Figure \ref{fig:diffPatterns}.
We clearly see the far-field diffraction signal that points outward from the center of the diffraction pattern (the center of the reciprocal space). The diffraction signal gets brighter in the longer exposed image of
Figure \ref{fig:diffPatterns}b) and c).


We filled the data behind the beamstop in the center of each long exposures
(Figure \ref{fig:diffPatterns}b) and c))
by the corresponding central part of the short exposure
(Figure \ref{fig:diffPatterns}a)).
The following definition was used to obtain a HDR diffraction pattern:
$
I^\textrm{HDR}_i = ( I^\textrm{noBS}_i  + I^\textrm{BS}_i \cdot \textrm{mask}_\textrm{BS}) \cdot  t_\textrm{BS} / \left(  t_\textrm{noBS} + t_\textrm{BS} \cdot \textrm{mask}_\textrm{BS}\right) .
\label{eq:IntensityScaling}
$
Here $i$ denotes the index of two corresponding diffraction patterns and $\textrm{mask}_\textrm{BS}$ is a binary mask that selects only the pixels outside the beam stop area. The HDR diffraction patterns were scaled to the long exposure time ($t_\textrm{BS1}$) with beam stop and the exposure time  $t_\textrm{noBS}$ without beam stop. Here only the relative exposure times are relevant. The combined data from
Figure \ref{fig:diffPatterns}a) and b) was combined in a subsequent step with a second beam stop mask and the longest exposure ($t_\textrm{BS2}$) of Figure \ref{fig:diffPatterns}c).

\begin{figure}[htbp]
\centering
\includegraphics[width=0.95\linewidth]{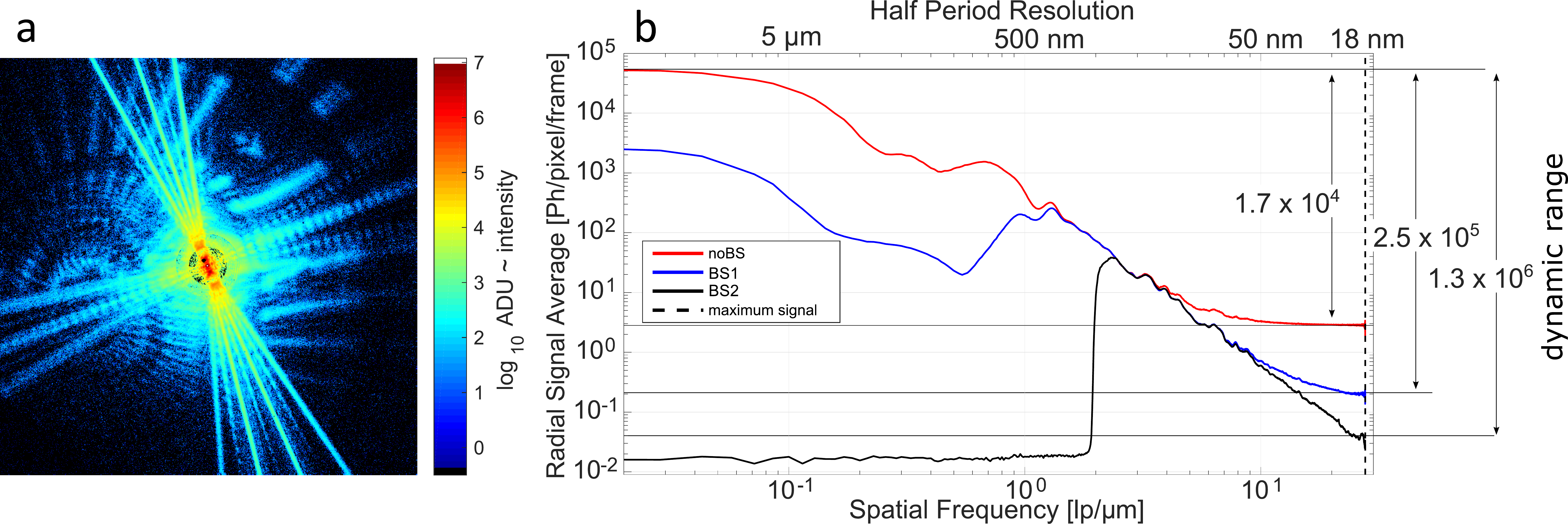}
\caption{ a) Merged Siemens star diffraction pattern of Figure \ref{fig:diffPatterns}. b) Power spectral density with (blue and black) and without beam stop (red) with indication of the dynamic range.}
\label{fig:PSD}
\end{figure}

A DR estimate of the measured diffraction signal can be obtained by evaluating its power spectral density (PSD) in photon units. This is the averaged signal over angle as a function of the distance to the center of the diffraction pattern. The PSD of all three data sets is shown in
Figure \ref{fig:PSD}.
The PSD analysis was applied on the average of all diffraction patterns with the same exposure.
The red curve shows the natural signal decay for increasing spatial frequencies measured in line-pairs per micro meter (\si{lp\per\micro\meter}). The noisy background dominates the signal beyond
\SI{10}{lp\per\micro\meter} (equal to \SI{50}{\nano\meter} half period resolution).
The blue curve is the diffraction signal weighted with the characteristic profile of the semi-transparent beam stop. The signal decays exponentially until the dark current noise dominates beyond
\SI{20}{lp\per\micro\meter}.
The black curve follows the form of the blue line except that the signal is dominating over noise till the maximum measured spatial frequency of
\SI{28}{lp\per\micro\meter}.
By scaling according to the relative exposures of the red ($t_\textrm{noBS} $), blue ($t_\textrm{BS1} $) and the black ($t_\textrm{BS2} $) PSD curve from different exposures we got an effective increase of the DR from
\num{1.7e4}
to
\num{1.3e6}.
Our measurement strategy has led to a gain in DR by a factor of
\num{76}.
As a consequence of the DR improvement we were able to measure diffraction signal till the corners (note factor $\sqrt{2}$ in the following equation) of the square detector. We cropped to $N_\textrm{px} = \SI{1920}{pixels}$ in each dimension to center the diffraction patterns. The detector pixel size was $\Delta_\textrm{D} = \SI{13.5}{\micro\meter}$ and the corresponding maximum half period resolution of
\SI{18}{\nano\meter}
was calculated with
$\Delta_\textrm{res} = \lambda z / \sqrt{2} N_\textrm{px} \Delta_\textrm{D} $.

\section{Ptychographic reconstruction from HDR data}
\begin{figure}[htbp]
\centering
\includegraphics[width=0.84\linewidth]{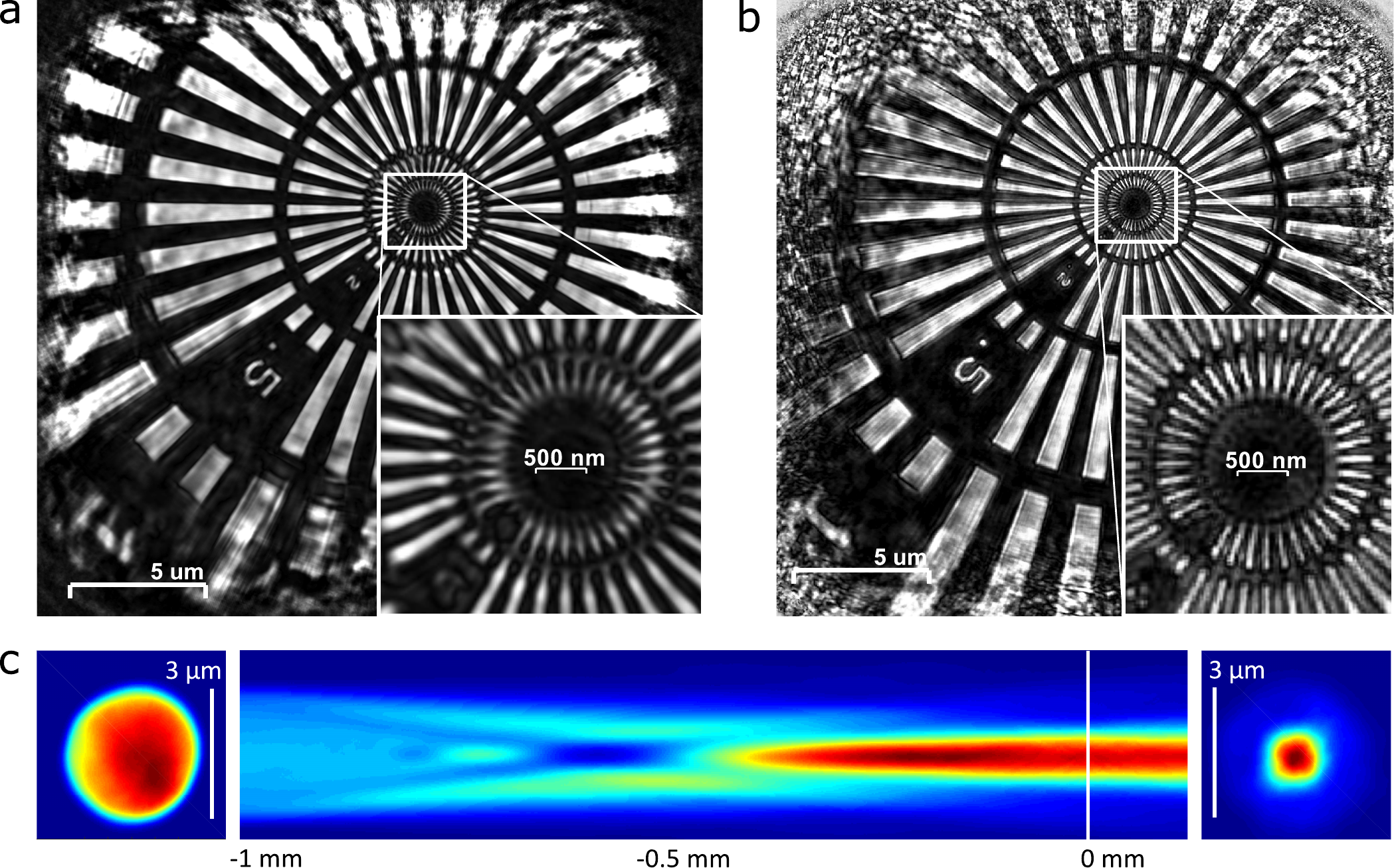}
\caption{Results of ptychographic reconstruction. Siemens star amplitude reconstruction from a) the single exposure data set and b) the HDR data set with improved quality for the \SI{50}{\nano\meter} structures. c) Reconstructed illumination function in the pinhole plane \SI{-1}{\milli\meter} (left) and propagated to the object plane at \SI{0}{\milli\meter} (right). }
\label{fig:reconstruction}
\end{figure}

The reconstruction was done with the ePIE algorithm and
\num{4000}
iterations on the single exposure data without beam stop
(see Figure \ref{fig:reconstruction}a)).
The relatively weak signal close to the detector edges shown in
Figure \ref{fig:diffPatterns}a)
led to a reduced reconstruction quality for the very fine
\SI{50}{\nano\meter}
structures in the center of the test pattern. To increase resolution we have combined the data with and without beam stop.
For the HDR ptychographic reconstruction we started with the merged data set. After \num{10} initial iterations we mask these values from the modulus constraint step in order to let the algorithm find the best estimate for the values inside the beam stop area.  This modification to the original ePIE algorithm can correct for scaling and position imperfections between the data sets used for the HDR merging. The corresponding reconstruction resulted in an improved image of our test sample shown in
Figure \ref{fig:reconstruction}b).
With the HDR data the
\SI{50}{\nano\meter}
structures are clearly resolved with two pixels per line.

Together with the object reconstruction we have obtained an illumination function at the sample plane. The retrieved complex valued illumination was propagated back to the pinhole in order to determine a precise distance between pinhole and sample. In
Figure \ref{fig:reconstruction} c)
the illumination in the pinhole and sample plane and a horizontal slice along the propagation distance is shown. At the pinhole plane the illumination takes the shape of the pinhole with
\SI{3}{\micro\meter}
in diameter. At the sample plane downstream of the pinhole
the illumination has a full width of half maximum of
\SI{1.25}{\micro\meter}.

\section{Conclusions}

We have presented a HDR ptychographic experiment in the water window photon energy range. Our HDR data were obtained by merging multiple increasing exposures into a single data set. To protect the detector from the highly intense X-ray beam we used beam stops. The measured DR was improved by roughly two orders of magnitude. The DR improvement led directly to an improvement of the reconstructed test object. Together with the object we obtained a high quality complex valued illumination profile.

Due to the radiation hardness of our test sample we do not suffer from radiation damage and the multiple exposure scheme is a good tool to obtain very high resolution diffraction patterns from a comparably low-dynamic-range detector. For more radiation sensitive biological objects we may restrict the number of exposures to two instead of three. Our HDR method can be applied without complications to cryo-fixated and hydrated biological samples to obtain high resolution phase and absorption contrast images.

\section{References}
\bibliographystyle{iopart-num_mod}
\bibliography{lit}

\end{document}